\documentclass[pdflatex,sn-mathphys-num]{sn-jnl}% Math and Physical Sciences Numbered Reference Style
\usepackage{graphicx}%
\usepackage{multirow}%
\usepackage{amsmath,amssymb,amsfonts}%
\usepackage{amsthm}%
\usepackage{mathrsfs}%
\usepackage[title]{appendix}%
\usepackage{xcolor}%
\usepackage{textcomp}%
\usepackage{manyfoot}%
\usepackage{booktabs}%
\usepackage{algorithm}%
\usepackage{algorithmicx}%
\usepackage{algpseudocode}%
\usepackage{listings}%

\usepackage{setspace}%Для изменения межстрочных расстояний в пределах абзаца для титульной страницы Suppl Inf

\usepackage{array}

\theoremstyle{thmstyleone}%
\theoremstyle{thmstyletwo}%

\theoremstyle{thmstylethree}%

\usepackage[T1,T2A]{fontenc}
\usepackage[utf8]{inputenc}
\usepackage[russian,english]{babel}

\begin{document}

\title[]{Sharp Coherence Crossover and Multimode Phase Control in Coupled Exciton-Polariton Condensates}

\author[1,2]{R.\,V.\,~Cherbunin}\email{r.cherbunin@rqc.ru}
\author[1,2]{A.\,D.\,~Liubomirov}
\author[2]{M.\,A.\,~Chukeev}
\author[3,2,4]{E.~S.~Sedov}\email{evgeny\_sedov@mail.ru}
\author[1,2,3,5]{A.~V.~Kavokin}

\affil[1]{\orgdiv{Russian Quantum Center}, \orgname{Skolkovo}, \orgaddress{\street{Bolshoi boulevard 30}, \city{Moscow}, \postcode{121205}, \country{Russia}}}

\affil[2]{\orgdiv{Spin Optics Laboratory}, \orgname{St. Petersburg State University}, \orgaddress{\street{Ulyanovskaya 1}, \city{St. Petersburg}, \postcode{198504}, \country{Russia}}}

\affil[3]{\orgdiv{Abrikosov Center for Theoretical Physics}, \orgname{Moscow Institute of Physics and Technology}, \orgaddress{\street{Institutskiy per. 9}, \city{Dolgoprudny}, \postcode{141701}, \state{Moscow Region}, \country{Russia}}}

\affil[4]{\orgname{Stoletov Vladimir State University}, \orgaddress{\street{Gorky str. 87}, \city{Vladimir}, \postcode{600000}, \country{Russia}}}

\affil[5]{\orgdiv{School of Physics}, \orgname{Westlake University}, \orgaddress{\street{Dunyu Road 600}, \city{Hangzhou}, \postcode{310030}, \country{China}}}



\abstract{
We investigate the formation of mutual coherence and deterministic phase control in a multimode exciton-polariton system realized in two optically induced traps with tunable intertrap coupling.
As the coupling channel between the traps is progressively opened, the mutual first-order coherence rises sharply from nearly zero to values close to unity, revealing a threshold-like crossover from weakly correlated to highly coherent dynamics.
We demonstrate deterministic multimode phase control with a short spatially localized nonresonant pulse, which allows the phase relations between selected polariton condensate modes to be set while keeping their amplitudes nearly unchanged.
The induced phase transformation is read out directly from time-resolved interferograms and realized for different sets of modes in both intermediate- and high-coherence regimes.
Within a reduced four-mode description, the control pulse implements an effective phase operator on the coherent multimode field with a fidelity exceeding 0.995.
The same approach can be extended to larger networks of coupled traps, offering a route to optical control over phase relations in increasingly complex multimode polariton states.
}

%%================================%%
%% Sample for structured abstract %%
%%================================%%

%\keywords{exciton-polaritons, spin-caloritronics, spin–orbit interaction, Seebeck effect, spin Seebeck effect, spin Nernst effect, longitudinal optical spin Nernst effect}



\maketitle
\date{\today}
{
%\color[rgb]{0,0.63,1} Light blue
%\color{red} Red
%\color[rgb]{1,0.52,0} Orange
%\color[rgb]{0,.64,0} Green
%\color[rgb]{1,.25,1} Pink
%\color[rgb]{0.58,.17,0.58} Purple
%\color[rgb]{0.5,0,0} Maroon
%\color[rgb]{0,1,1} Cyan
%\color[rgb]{0.6,0.3,0} Brown
%\color[rgb]{0,0,0.5} Navy
%\color[rgb]{0,0.5,0.5} Teal
%\color[rgb]{0.5,0.5,0} Olive
%\color[rgb]{0.75,0.75,0.75} Silver
%\color[rgb]{0.29,0,0.51} Indigo
%\color[rgb]{0.75,1,0} Lime
}

%SSSSSSSSSSSSSSSSSS
%EEEEEEEEEEEEEEEEEE
%CCCCCCCCCCCCCCCCCC
\section*{Introduction}

Among driven bosonic systems, exciton-polariton condensates offer a distinctive combination of macroscopic coherence, reconfigurability, and the possibility of probing the condensate state directly in standard optical experiments across readily accessible spatial and temporal scales.
Exciton-polaritons are bosonic quasiparticles formed through the strong coupling of cavity photons and quantum-well excitons in semiconductor optical microcavities~\cite{kavokinBook2017}.
Their hybrid light-matter nature combines the optical accessibility inherited from photons with the appreciable interactions and nonlinearities provided by excitons.
The photonic component continuously escapes through the cavity mirrors, carrying information about the condensate density, coherence, and spatial phase that can be obtained through routine interferometric measurements~\cite{PhysRevLett99126403,Nature443409,ACSPhot71163,SciRep134607}.
The same radiative loss makes the condensate intrinsically dissipative and requires its continuous replenishment from an external source.

Under nonresonant optical excitation~\cite{Nature443409,ACSPhot71163,SciRep134607}, with the pump photon energy well above the polariton energy, this replenishment is mediated by an incoherent reservoir of high-energy excitons created by the pump.
The relaxation of reservoir excitons feeds the macroscopically occupied polariton state, thereby compensating for the losses.
At the same time, the reservoir repels polaritons and produces a local energy blueshift that acts as an effective potential for polaritons.
A spatially structured pump therefore shapes both the gain profile and the reservoir-induced potential landscape, which together select the populated condensate modes and govern their spatial overlap and mutual coherence.
The reservoir-induced potential can also complement static structural confinement, enabling the formation of condensate states, including those with controllable circulation~\cite{ACSPhot71163,SciRep134607,PhysRevB97195149,PhysRevResearch3013072,PhysRevApplied22054031,Optica12991}.
In addition, reservoir fluctuations may act as a source of dephasing and thereby affect the formation of long-lived coherent polariton states~\cite{CherbuninIntratrapCoher2026}.
The interplay of gain, confinement, and dephasing becomes particularly important in optically induced traps, where all three are determined by the imposed pump geometry.

Optical traps provide particularly flexible control over polariton condensates.
A shaped pump can surround a region of the microcavity with a repulsive excitonic barrier, allowing polaritons to condense away from the region of highest reservoir density.
Ring-shaped excitation has been used to confine condensates in optically generated two-dimensional potentials~\cite{PhysRevB88041308,PhysRevB97235303,PhysRevB101245309}, while more elaborate pump geometries have shown that sculpting the reservoir can reconfigure both the trapping landscape and condensate dynamics~\cite{NatPhys8190,PhysRevB103235313,PhysRevLett113200404, PhysRevLett128117401,PhysRevResearch3013099}.
Annular traps have supported the spontaneous formation of vortices~\cite{PhysRevB101245309,SciRep1412953}, and chiral deformations of the same basic geometry have transferred orbital angular momentum to the condensate~\cite{PhysRevLett113200404}.
Beyond static geometries, time-dependent optical excitation has been used to create rotating traps, enabling controlled vortex formation and coupling between the orbital and polarization dynamics of the condensate~\cite{SciAdv9eadd1299,PhysRevB108045301,PhysRevB108155301,PhysRevResearch6013261}.
The discrete spectrum of annular traps has also been resolved directly, revealing spatially quantized condensate modes whose energies and wavefunctions are controlled by the trap geometry~\cite{PhysRevB107045302}.
Spatially separating the condensate from the reservoir also has a marked effect on its coherence.
As reported in~\cite{PhysRevB103235313}, the coherence time of an optically confined state can exceed one nanosecond, more than an order of magnitude longer than in an untrapped condensate.
Optical trapping thus provides both a reconfigurable modal structure and favourable conditions for coherent polariton dynamics.

When the potential landscape supports more than one condensate, the relative phase between the condensates becomes central to their collective behaviour.
Spatially separated yet coupled polariton condensates have displayed zero- and $\pi$-phase ordering, Josephson oscillations, and self-trapping, all arising from a balance between tunnelling, interactions, gain, and dissipation~\cite{Nature450529,PhysRevLett105120403,NatPhys9275}.
Coupling between excited states of optically trapped condensates has also been proposed to realize polariton molecules with $\sigma$- and $\pi$-like bonding configurations~\cite{PhysRevB103115309,Optica1393}.
In optically generated condensate multiplets, the selected relative phase was found to vary nontrivially with the distance between the excitation spots, as the phase accumulated by outflowing polaritons modifies the effective coupling~\cite{PhysRevX6031032}.
At larger distances, the propagation time itself becomes relevant and can give rise to delayed coupling dynamics~\cite{CommPhys32}.
Recent experiments have demonstrated further mechanisms of interaction, ranging from dimers with geometrically tunable ballistic and evanescent coupling~\cite{NatCommun169794} to phase locking between otherwise isolated condensates mediated by optical feedback from an external mirror~\cite{liang2026mirrormediatedlongrangecouplingrobust}.
Despite this diversity of coupling mechanisms, a central question for controlled phase manipulation is how mutual coherence develops as two separately trapped condensates are continuously brought into interaction.

Once mutual coherence has been established, the spatial modes supported by the coupled system themselves become available for coherent manipulation.
Structured optical potentials have already been used to control orbital and spin degrees of freedom, for example through the spin-dependent splitting of a propagating polariton fluid~\cite{PhysRevB97235303}.
In annular traps, the circulation direction can be selected stochastically, while a weak displaced control pulse breaks the symmetry and selects it deterministically~\cite{SciRep1412953}.
A related experiment demonstrated persistent circulation initiated on demand by a short optical pulse and lasting for many revolutions without further stimulation~\cite{Optica12991}.
In an elliptic trap, a localized nonresonant pulse transiently deformed the potential and changed the frequency, amplitude, and phase of beating between two size-quantized polariton  modes~\cite{MDPIOptics653}.
These results show that optical perturbations can act on the phase dynamics of confined condensates without resonantly imposing the phase of the driving field.
Extending this principle from a multimode condensate confined to a single trap to one distributed across coupled traps calls for a reproducible operation on the phase relations among its modes, together with a measurement capable of resolving the induced transformation.

In this paper, we study a multimode exciton-polariton system in a double-well potential formed by two optically induced traps with a tunable center-to-center separation.
We find that the mutual first-order coherence between the two traps undergoes a sharp crossover, remaining close to zero for well-separated traps and rising rapidly to values near unity as the traps are brought together.
We demonstrate deterministic control over the phase relations among populated polariton condensate modes in both intermediate- and high-coherence regimes by applying a spatially localized nonresonant optical pulse.
The action of the control pulse on the coherent multimode polariton field can be described with high fidelity by an effective phase operator, while the corresponding phase shifts are directly resolved from interferometric measurements.
At intermediate coherence, this phase control remains effective, while a two-component model combining coherent and statistically averaged contributions accurately reproduces the measured interference pattern.
Our results connect the development of mutual coherence with deterministic and directly readable phase control in reconfigurable multimode polariton condensates in coupled optical traps.

%SSSSSSSSSSSSSSSSSS
%EEEEEEEEEEEEEEEEEE
%CCCCCCCCCCCCCCCCCC
\section*{Results}

%SSSSSSSSSSSSSSSSSS
%SSSSSSSSSSSSSSSSSS
%EEEEEEEEEEEEEEEEEE
%CCCCCCCCCCCCCCCCCC
\subsection*{Experimental system and coherence measurement}

%FFFFFFFFFFFFFFFFFFFFFFFF
%IIIIIIIIIIIIIIIIIIIIIIII
%GGGGGGGGGGGGGGGGGGGGGGGG
\begin{figure*}
\includegraphics[width=\textwidth]{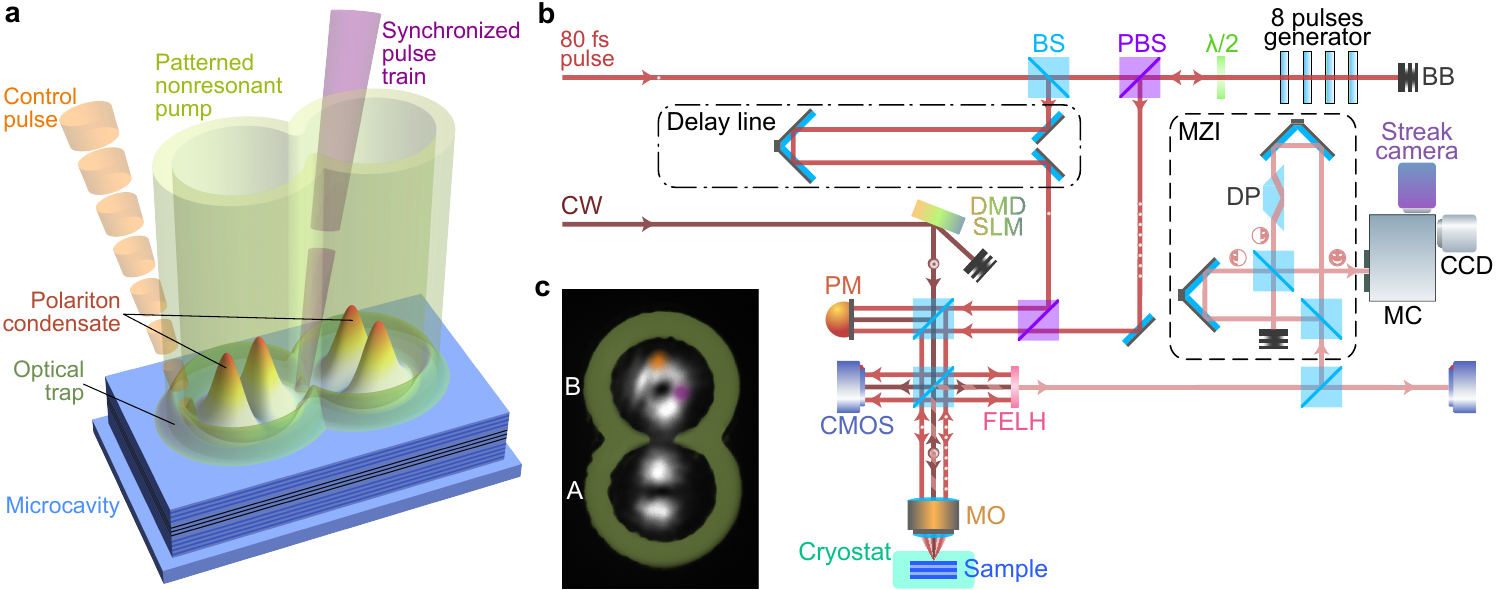}
\caption{\label{FIG_Scheme}
\textbf{Control of multimode polariton condensates in coupled optical traps.}
\textbf{a},~Schematic of two adjacent optical traps formed in a semiconductor microcavity by a patterned nonresonant CW pump.
A synchronized pulse train reinforces the multimode condensate oscillations, while a spatially localized control pulse manipulates their phase evolution.
\textbf{b}, Schematic of the optical setup comprising a CW excitation branch with an SLM for forming the coupled traps, a pulsed branch with an adjustable delay line and a pulse generator for synchronized excitation and phase control, as well as interferometric, time-integrated, and time-resolved detection channels.
Further details of the optical setup are provided in Methods.
\textbf{c}, Representative time-integrated real-space PL of the polariton condensates.
The trap profile reconstructed from the below-threshold PL is superimposed in green, while the coloured spots indicate the positions of the synchronized pulse train (purple) and the control pulse (orange).
The two traps are labelled $A$ and $B$.
}
\end{figure*}

The system studied in this work is formed by two adjacent optically induced traps in a planar semiconductor microcavity, as schematically shown in Fig.~\ref{FIG_Scheme}a.
Each trap is defined by a ring-shaped nonresonant laser pump beam that creates a reservoir of photoexcited excitons.
The reservoir forms a repulsive potential barrier that confines polaritons within the central region.
When the two traps overlap, the pump profile is modified to introduce an intertrap channel in the barrier within the overlap region.

In the experiment, we studied a planar AlGaAs-based $3\lambda/2$ microcavity.
The optical excitation and detection arrangement is shown in Fig.~\ref{FIG_Scheme}b.
A continuous-wave (CW) single-mode semiconductor laser was used to create the optical traps.
The pump beam was patterned with a spatial light modulator (SLM), allowing the shape and relative position of the two traps and the geometry of the intertrap channel to be varied without changing the remaining excitation conditions.
A representative time-integrated real-space photoluminescence (PL) image of the polariton condensates is shown in Fig.~\ref{FIG_Scheme}c, overlaid with the trap profile reconstructed from the incoherent PL recorded below the condensation threshold.

Owing to the driven-dissipative nature of polariton condensates, several modes with sufficient gain to compensate their losses can coexist in the trapping potential.
The  coherent evolution of these modes gives rise to oscillations of the PL intensity governed by the frequency differences between the  modes.
To deliberately reinforce these oscillations, we supplemented the CW pump with a synchronized sequence of eight nonresonant femtosecond pulses generated in the pulsed branch of the setup schematically shown in Fig.~\ref{FIG_Scheme}b.
The interval between successive pulses was adjusted for each trap configuration to match the observed oscillation period.

The emitted light was analysed through the complementary time-resolved and interferometric detection channels shown in Fig.~\ref{FIG_Scheme}b.
Streak-camera measurements resolved the temporal oscillations of the PL intensity, whereas a Mach--Zehnder interferometer (MZI) provided access to the spatial coherence and phase evolution of the polariton field.
To deliberately control the relative phases of the populated states, we applied an additional spatially localized nonresonant pulse to a selected region of one of the traps.
Its position and arrival time were chosen to perturb the ongoing phase evolution in a controlled manner.
The positions of the synchronized pulse train and the control pulse on the sample are indicated in Fig.~\ref{FIG_Scheme}c.
Further details of the excitation and detection procedures are provided in Methods.

%SSSSSSSSSSSSSSSSSS
%SSSSSSSSSSSSSSSSSS
%EEEEEEEEEEEEEEEEEE
%CCCCCCCCCCCCCCCCCC
\subsection*{Sharp crossover of mutual coherence}

Optically induced traps can support a discrete spectrum of spatially quantized polariton states, enabling rich multimode condensate dynamics~\cite{PhysRevB107045302}.
Rich as this physics may be, it remains confined to a single trap and is therefore of only limited interest once the focus shifts from local mode structure to collective behaviour.
The more intriguing regime emerges when two traps are brought into interaction, allowing multimode phase correlations to develop across both traps.
Our experiment therefore traces how the system evolves from nearly independent polariton fields in the two traps to coherent dynamics spanning both traps as the intertrap coupling is varied in a controlled manner.
We quantify this crossover through the mutual first-order coherence between the two traps.

To measure this coherence, the real-space PL image was split between the two arms of the MZI.
In one arm, the image was reflected about the horizontal axis, $y\rightarrow -y$, to superimpose the two trap regions.
The images were recombined at a small angle, introducing a carrier wave vector component $q_{y}$, with the complex first-order coherence encoded in the interference fringes.
Varying the delay $\tau$ between the arms yielded the spatially resolved mutual coherence $g^{(1)}(\tau,\mathbf{r})$.

%FFFFFFFFFFFFFFFFFFFFFFFF
%IIIIIIIIIIIIIIIIIIIIIIII
%GGGGGGGGGGGGGGGGGGGGGGGG
\begin{figure*}
\includegraphics[width=\textwidth]{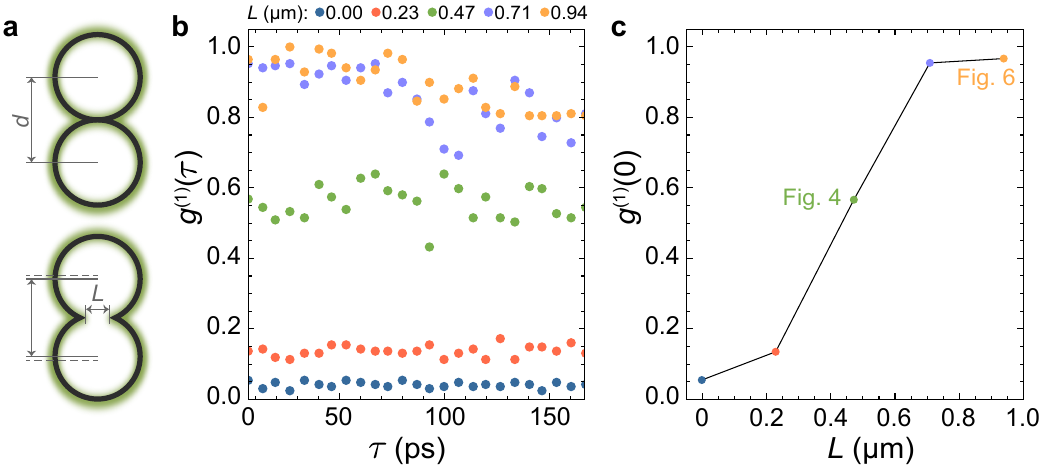}
\caption{\label{FIG_Coherence}
\textbf{Sharp crossover of mutual coherence in coupled polariton condensates.}
\textbf{a},~Definition of the control parameter $L$ used to tune the intertrap coupling.
The black contours show the binary ring patterns imposed on the SLM for two center-to-center separations $d$.
When the two rings intersect, the segment of the pattern between their intersection points is removed, and the separation between these points defines the channel width $L$.
The green shading schematically indicates the corresponding smooth pump profile on the sample.
\textbf{b}, Delay dependences of the first-order coherence $g^{(1)}(\tau)$ for five channel widths $L$.
\textbf{c}, Zero-delay coherence $g^{(1)}(0)$ as a function of channel width $L$.
The line connecting the points is a guide to the eye.
The points labelled Fig.~4 and Fig.~6 indicate the two trap configurations selected for the phase-control measurements discussed below.
}
\end{figure*}

The intertrap coupling was tuned through the width $L$ of a channel formed in the pump-induced reservoir barrier.
As illustrated in Fig.~\ref{FIG_Coherence}a, each trap was defined by a binary ring pattern on the SLM.
When the two rings intersected, the segment of the pattern between their intersection points was removed, and the separation between these points defined $L$.
Although the sharp boundaries of the binary pattern were smoothed in the projected pump profile, $L$ remained a directly controlled geometrical parameter.

The delay dependences $g^{(1)}(\tau)\equiv g^{(1)}(\tau,\mathbf{r}^{\ast}_L)$, evaluated at the point $\mathbf{r}^{\ast}_L$ of maximum intensity overlap between the superimposed images, are shown for different $L$ in Fig.~\ref{FIG_Coherence}b.
For a closed or narrow channel, $g^{(1)}(\tau)$ remains close to zero throughout the measured delay range, with no resolvable delay dependence.
As the channel is widened, a pronounced coherence peak develops at zero delay and decays with increasing $\tau$.

The central result of these measurements becomes apparent from the dependence of the zero-delay coherence $g^{(1)}(0)$ on $L$, shown in Fig.~\ref{FIG_Coherence}c.
The coherence initially remains close to zero and then rises rapidly over a narrow range of $L$, bringing the system into a high-coherence regime.
Although the sampling of $L$ is insufficient to resolve the detailed shape of the crossover or assign a specific crossover channel width, the data clearly reveal a sharp, threshold-like onset of coherence between the two traps.
At the same time, the observed PL oscillations mentioned above, arising from beating between spatial modes with different frequencies, reveal the multimode character of the condensate dynamics.
The coexistence of intertrap coherence and multimode dynamics motivates the complementary reduced descriptions introduced in the following section.

%SSSSSSSSSSSSSSSSSS
%SSSSSSSSSSSSSSSSSS
%EEEEEEEEEEEEEEEEEE
%CCCCCCCCCCCCCCCCCC
\subsection*{Reduced four-mode descriptions}

Under the experimental conditions, the balance of gain and losses selects a limited set of spatial modes that dominate the observed polariton condensate dynamics.
A minimal description of each isolated trap retains two such modes, yielding a four-mode subspace for the coupled system.
The appropriate four-mode description depends on the degree of mutual coherence between the traps.
For weakly correlated polariton fields, the relevant modes remain associated with the individual traps, whereas in the coherent regime they extend across both traps.
These reduced descriptions do not aim to reproduce the complete driven-dissipative dynamics, but retain the modal content and correlations needed to capture the observed multimode evolution.

\begin{figure*}
\includegraphics[width=0.8\textwidth]{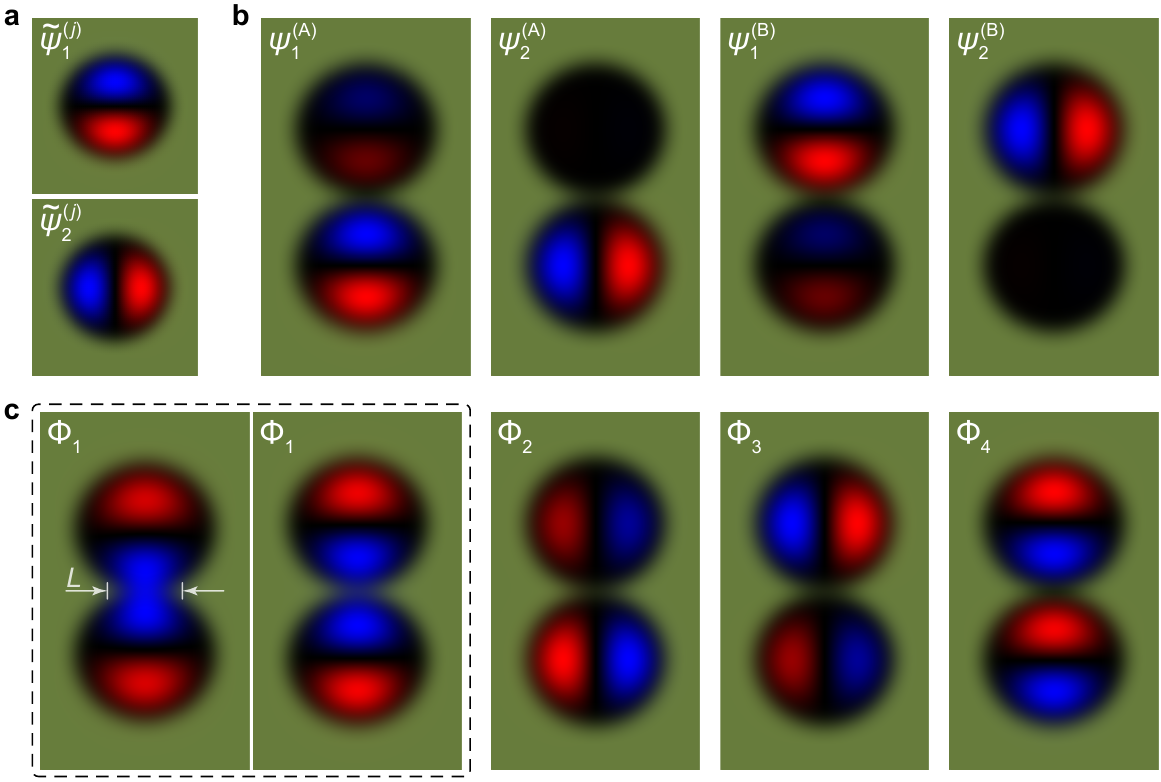}
\caption{\label{FIG_Eigs}
\textbf{Construction of a basis of states for describing polariton condensates in coupled optical traps.}
\textbf{a}, Representative orthogonal dipole-like states $\tilde{\psi}_{1}^{(j)}(\mathbf{r})$ and $\tilde{\psi}_{2}^{(j)}(\mathbf{r})$ of an isolated elliptic trap.
\textbf{b}, Four single-trap states $\psi_{1}^{(A)}(\mathbf{r})$, $\psi_{2}^{(A)}(\mathbf{r})$, $\psi_{1}^{(B)}(\mathbf{r})$, and $\psi_{2}^{(B)}(\mathbf{r})$ forming an orthonormal prebasis for the two-particle model.
\textbf{c}, Four eigenstates $\Phi_{1}(\mathbf{r})$--$\Phi_{4}(\mathbf{r})$ of the double-trap potential used in the coherent-field description.
Two realizations of $\Phi_{1}(\mathbf{r})$ within the dashed frame correspond to different intertrap channel widths $L$.
Red and blue denote opposite phases of the wavefunction.
The potential profile is superimposed on each spatial map in green.
}
\end{figure*}

We first obtain the single-trap modes from separate conservative descriptions of the polariton condensate field confined by the model potentials $V_{j}(\mathbf{r})$, where $j=A,B$ labels the two traps.
The explicit form of the potentials is discussed in Methods.
For each trap $j$, the relevant modes are obtained from the stationary eigenvalue problem
\begin{equation}
\hat{H}_{\mathrm{sing}}^{(j)}(\mathbf{r})
\tilde{\psi}_{n}^{(j)}(\mathbf{r})
=
\hbar\tilde{\omega}_{n}^{(j)}
\tilde{\psi}_{n}^{(j)}(\mathbf{r}),
\qquad
n=1,2,
\label{EqSingleTrapEigenproblemMain}
\end{equation}
where $\hat{H}_{\mathrm{sing}}^{(j)}(\mathbf{r}) = -\hbar^{2}\nabla^{2}/2m^{*} + V_{j}(\mathbf{r})$ is the Hamiltonian of the isolated trap $j$ and $m^{*}$ is the effective cavity polariton mass.
For each trap, we retain two dipole-like modes, $\tilde{\psi}_{1}^{(j)}(\mathbf{r})$ and $\tilde{\psi}_{2}^{(j)}(\mathbf{r})$.
Representative spatial profiles of these modes are shown schematically in Fig.~\ref{FIG_Eigs}a.

%FFFFFFFFFFFFFFFFFFFFFFFF
%IIIIIIIIIIIIIIIIIIIIIIII
%GGGGGGGGGGGGGGGGGGGGGGGG

%SSSSSSSSSSSSSSSSSS
%SSSSSSSSSSSSSSSSSS
%SSSSSSSSSSSSSSSSSS
%EEEEEEEEEEEEEEEEEE
%CCCCCCCCCCCCCCCCCC
\subsubsection*{Two-particle description}

We now turn to the coupled-trap system.
In the regime of weak intertrap correlations, the condensate field in each trap largely retains the spatial character of the corresponding single-trap modes.
The reduced four-dimensional prebasis is therefore constructed from the two retained modes of each isolated trap,
$ \tilde{\boldsymbol{\psi}} =
(
\tilde{\psi}_{1}^{(A)},
\tilde{\psi}_{2}^{(A)},
\tilde{\psi}_{1}^{(B)},
\tilde{\psi}_{2}^{(B)}
)^{\mathrm{T}}$.

Once the traps overlap, however, the four prebasis functions are no longer mutually orthogonal.
We therefore construct an orthonormal set by symmetric orthogonalization.
Introducing the composite indices $\alpha=(a,j)$ and $\alpha'=(a',j')$, we define the overlap matrix
$\mathcal{M}_{\alpha\alpha'} =
\int [\tilde{\psi}_{a}^{(j)}(\mathbf{r})]^{*}
\tilde{\psi}_{a'}^{(j')}(\mathbf{r})
\,d^{2}\mathbf{r}$.
The orthonormal basis is then obtained as
\begin{equation}
\boldsymbol{\psi}
= (\psi_{1}^{(A)},
\psi_{2}^{(A)},
\psi_{1}^{(B)},
\psi_{2}^{(B)}
)^{\mathrm{T}}
= \mathcal{M}^{-1/2}\tilde{\boldsymbol{\psi}}.
\end{equation}
Representative spatial profiles of the resulting basis states are shown schematically in Fig.~\ref{FIG_Eigs}b.

Following~\cite{KudlisDensMatrFormal}, we assign independent two-dimensional coordinates $\mathbf{r}_{A}$ and $\mathbf{r}_{B}$ to the condensate fields associated with traps $A$ and $B$, respectively.
Each coordinate spans the full plane of the microcavity, while the localized basis functions determine where the corresponding field is predominantly concentrated.
The joint state is then represented by the two-coordinate wave function
\begin{equation}
\Psi(t,\mathbf{r}_{A},\mathbf{r}_{B})
= \sum_{a,b=1}^{2}
C_{\mu}(t) \psi_{a}^{(A)}(\mathbf{r}_{A}) \psi_{b}^{(B)}(\mathbf{r}_{B}),
\qquad  \mu=2(a-1)+b.
\label{EqTwoParticleWaveFunction}
\end{equation}
Here, $a$ and $b$ label the two basis states associated with traps $A$ and $B$, respectively, while the single index $\mu=1,\ldots,4$ enumerates the resulting four product states and $C_{\mu}(t)$ are their complex amplitudes.

In our experiments, the polariton condensate dynamics was probed through two observables, the PL intensity and self-interference, both determined by bilinear combinations of the emitted optical field.
To reconstruct these observables within the model, we introduce the two-time correlation matrix $\varrho(t,t')=[\varrho_{\mu\mu'}(t,t')]$, with elements $\varrho_{\mu\mu'}(t,t')=C_{\mu}^{*}(t)C_{\mu'}(t')$.
This matrix describes correlations between the joint product states of the two-trap system and therefore does not directly provide the correlations of the fields associated with the individual traps.
Using the correspondence $\mu\leftrightarrow(a,b)$, these correlations are recovered through the reduced matrices
\begin{equation}
\rho_{aa'}^{(A)}(t,t') = \sum_{b=1}^{2} \varrho_{ab,a'b}(t,t'), \qquad 
\rho_{bb'}^{(B)}(t,t') = \sum_{a=1}^{2} \varrho_{ab,ab'}(t,t').
\end{equation}

To recover the spatial dependence of the observables, we return from the reduced mode space to the plane of the microcavity and define the first-order correlation function for each trap as
\begin{equation}
\Gamma_{j}(t,\mathbf{r};t',\mathbf{r}')
=
\sum_{n,n'=1}^{2}
\rho_{nn'}^{(j)}(t,t')
\left[\psi_{n}^{(j)}(\mathbf{r})\right]^{*}
\psi_{n'}^{(j)}(\mathbf{r}'),
\qquad
j=A,B.
\end{equation}
Within the two-particle description, the correlation function of the coupled system is reconstructed as $\Gamma_{\mathrm{2p}}(t,\mathbf{r};t',\mathbf{r}')
=
\sum_{j=A,B}\Gamma_{j}(t,\mathbf{r};t',\mathbf{r}')$.

The model counterpart of the measured PL intensity is the polariton condensate density, obtained by evaluating the correlation function at equal times and positions,
$I_{\mathrm{2p}}(t,\mathbf{r})
= \Gamma_{\mathrm{2p}}(t,\mathbf{r};t,\mathbf{r})$.
The corresponding interference pattern is obtained by superimposing the condensate field with its temporally delayed and spatially transformed copy. Introducing the  space-time arguments
$\xi_{0}=(t,\mathbf{r})$ and $\xi_{1}=(t+\tau,\hat{\mathcal{T}}\mathbf{r})$,
we write
$J_{\mathrm{2p}}(t,\mathbf{r};\tau)
= \sum_{\iota,\iota'=0}^{1}
e^{i(\iota'-\iota)\mathbf{q}\cdot\mathbf{r}}\,
\Gamma_{\mathrm{2p}}(\xi_{\iota};\xi_{\iota'})$.
Here, $\tau$ is the temporal delay, $\hat{\mathcal{T}}$ specifies the spatial transformation applied to one of the interfering fields, and $\mathbf{q}$ is the carrier wave vector.

%SSSSSSSSSSSSSSSSSS
%SSSSSSSSSSSSSSSSSS
%SSSSSSSSSSSSSSSSSS
%EEEEEEEEEEEEEEEEEE
%CCCCCCCCCCCCCCCCCC
\subsubsection*{Coherent-field description}

In the high-coherence regime, the condensate dynamics is naturally represented by a single field extending across both traps.
The reduced four-mode subspace is therefore constructed from four selected solutions of the eigenvalue problem for the coupled double-trap potential $W(\mathbf{r})$,
\begin{equation}
\hat{H}_{0}(\mathbf{r}) \Phi_{\nu}(\mathbf{r})
= \hbar\omega_{\nu} \Phi_{\nu}(\mathbf{r}),
\qquad \nu=1,\ldots,4,
\label{EqCoherentEigenproblem}
\end{equation}
where $\hat{H}_{0}(\mathbf{r}) = -\hbar^{2}\nabla^{2}/2m^{*} + W(\mathbf{r})$ is the stationary double-trap Hamiltonian and $\omega_{\nu}$ are the corresponding eigenfrequencies.
The explicit form of $W(\mathbf{r})$ is discussed in Methods.
The retained states extend across both traps and form an orthonormal basis for the coherent-field description.
Representative spatial profiles of these states are shown schematically in Fig.~\ref{FIG_Eigs}c.

The condensate field is expanded in this basis as
\begin{equation}
\Phi(t,\mathbf{r})
=
\sum_{\nu=1}^{4}
u_{\nu}(t)\Phi_{\nu}(\mathbf{r}),
\end{equation}
where $u_{\nu}(t)$ are the complex amplitudes of the retained states.
We introduce the two-time correlation matrix $\mathcal{G}(t,t')=[\mathcal{G}_{\nu\nu'}(t,t')]$ with $\mathcal{G}_{\nu\nu'}(t,t')=u_{\nu}^{*}(t)u_{\nu'}(t')$.
The first-order correlation function in the microcavity plane is then
\begin{equation}
\Gamma_{\mathrm{coh}}(t,\mathbf{r};t',\mathbf{r}')
=
\sum_{\nu,\nu'=1}^{4}
\mathcal{G}_{\nu\nu'}(t,t')
\left[\Phi_{\nu}(\mathbf{r})\right]^{*}
\Phi_{\nu'}(\mathbf{r}').
\label{EqCoherentCorrelationFunction}
\end{equation}

The polariton condensate density is obtained by evaluating the correlation function at equal times and positions,
$I_{\mathrm{coh}}(t,\mathbf{r})
=
\Gamma_{\mathrm{coh}}(t,\mathbf{r};t,\mathbf{r})$.
Using the same space-time arguments $\xi_{0,1}$ and carrier wave vector $\mathbf{q}$ as in the two-particle description, the corresponding interference pattern is written as
$J_{\mathrm{coh}}(t,\mathbf{r};\tau)
=
\sum_{\iota,\iota'=0}^{1}
e^{i(\iota'-\iota)\mathbf{q}\cdot\mathbf{r}} \,
\Gamma_{\mathrm{coh}}(\xi_{\iota};\xi_{\iota'})$.

%SSSSSSSSSSSSSSSSSS
%SSSSSSSSSSSSSSSSSS
%EEEEEEEEEEEEEEEEEE
%CCCCCCCCCCCCCCCCCC
\subsection*{Pulse-induced phase dynamics}

For a given trap configuration, the experimental excitation conditions select a reproducible set of polariton states that participate in the multimode condensate dynamics.
As mutual coherence develops between the traps, the phase relations among these states become experimentally accessible through the spatial and temporal evolution of the condensate field.
We next examine whether these phase relations can be brought under external optical control without substantially changing the set of participating states.
We address this question for the two trap configurations marked in Fig.~\ref{FIG_Coherence}c, representing the intermediate- and high-coherence regimes.
For each configuration, we trace the phase evolution by interfering the condensate emission with its temporally delayed copy reflected about the horizontal axis.
To manipulate this evolution, we apply an additional spatially localized nonresonant laser pulse to a selected region of the coupled-trap system.

%SSSSSSSSSSSSSSSSSS
%SSSSSSSSSSSSSSSSSS
%SSSSSSSSSSSSSSSSSS
%EEEEEEEEEEEEEEEEEE
%CCCCCCCCCCCCCCCCCC
\subsubsection*{Phase control in the intermediate-coherence regime}

\begin{figure*}
\includegraphics[width=0.75\textwidth]{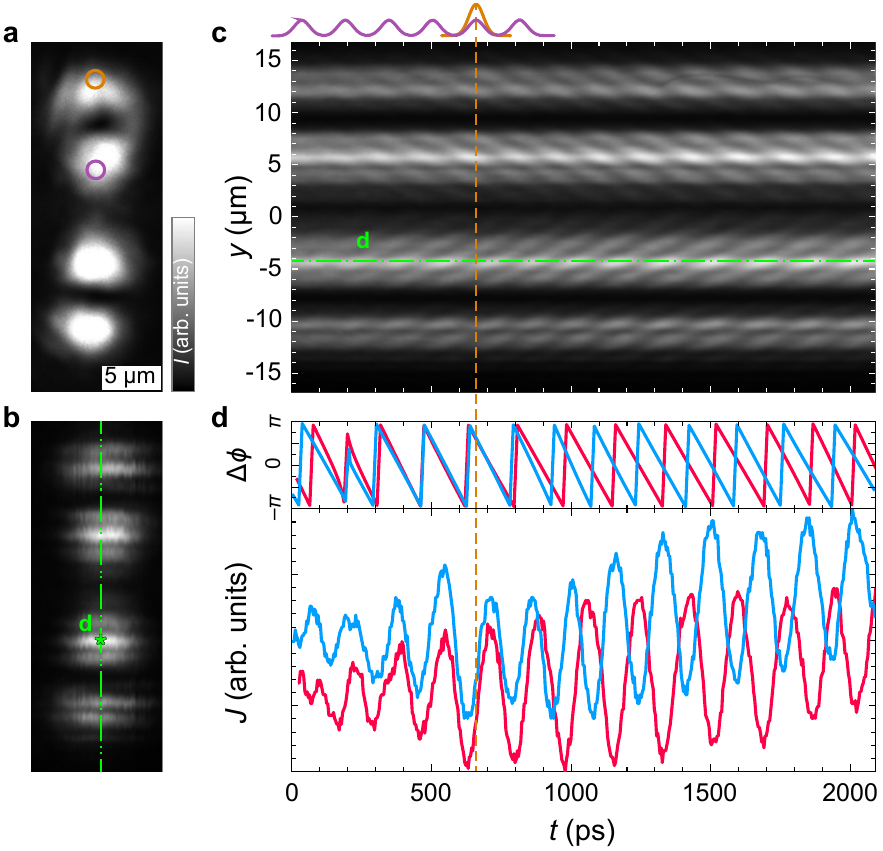}
\caption{\label{FIG_HalfQuant}
\textbf{Pulse-induced phase control in the intermediate-coherence regime.}
\textbf{a}, Time-integrated real-space PL of the polariton condensates.
The purple and orange circles mark the synchronized pulse train and the spatially localized control pulse, respectively.
\textbf{b}, Representative interferogram obtained by superimposing two real-space images of the condensate PL, one reflected about the horizontal axis.
\textbf{c}, Space--time evolution of the interferogram along the section indicated by the green dash-dotted line in \textbf{b}, measured in the presence of the control pulse.
The purple and orange waveforms above the panel indicate the synchronized pulse train and the control pulse, respectively.
The vertical orange dashed line marks the arrival time of the control pulse.
\textbf{d}, Interferometric signal $J(t)$ at the position marked by the green star in \textbf{b} and the green dash-dotted line in \textbf{c} (lower panel), together with the phase $\Delta\phi(t)$ of the corresponding oscillations (upper panel).
Red curves correspond to the reference measurement without the control pulse, whereas blue curves correspond to the otherwise identical measurement with the control pulse.
}
\end{figure*}

We first examine a trap configuration selected from the rising part of the coherence dependence in Fig.~\ref{FIG_Coherence}c.
Here, mutual coherence between the traps has already developed, while remaining below the plateau characteristic of the high-coherence regime.
Figure~\ref{FIG_HalfQuant}a shows the time-integrated PL of the polariton condensate for this configuration.
The spatial distribution of the emission indicates that the dynamics is dominated by states retaining the character of vertically oriented dipole-like modes of the individual traps.
A representative interferogram is shown in Fig.~\ref{FIG_HalfQuant}b.
Its predominantly horizontal fringes are set by the carrier wave-vector component $q_y$ introduced by the inclination of the interfering beams.

To resolve the temporal evolution of the interference pattern, we follow the interferogram along the $y$-directed section at $x=0$ indicated by the dash-dotted line in Fig.~\ref{FIG_HalfQuant}b.
This section runs along the axis connecting the traps and crosses the regions of strongest emission from the participating states.
The resulting space--time distribution in Fig.~\ref{FIG_HalfQuant}c exhibits pronounced oscillations of the PL intensity.
The synchronized pulse train is indicated schematically by the grey waveform above the panel.
The pulse spacing was matched to the observed oscillation period to reinforce the oscillatory dynamics.
The orange waveform marks the temporal profile and arrival time of the control pulse, whose spatial position within the trap is indicated by the orange circle in Fig.~\ref{FIG_HalfQuant}a.
The well-defined inclined fringes in the $(t,y)$ plane reflect the coherent phase evolution associated with beating between states of different frequencies~\cite{PhysRevLett105120403}.

%FFFFFFFFFFFFFFFFFFFFFFFF
%IIIIIIIIIIIIIIIIIIIIIIII
%GGGGGGGGGGGGGGGGGGGGGGGG

The effect of the control pulse is examined at the point in the interferogram marked by the green star in Fig.~\ref{FIG_HalfQuant}b.
In the space--time map in Fig.~\ref{FIG_HalfQuant}c, this fixed spatial position corresponds to the green dash-dotted line.
Figure~\ref{FIG_HalfQuant}d shows the interferometric signal $J(t)$ at this point (lower panel) and the phase $\Delta\phi(t)$ of the corresponding oscillations (upper panel).
The red curves show the reference measurement without the control pulse, whereas the blue curves were recorded under otherwise identical excitation conditions with the control pulse arriving at the time marked by the vertical orange dashed line.
Before the pulse arrives, the two interferometric traces oscillate in phase.

The phase shift develops only after a finite delay following the arrival of the control pulse and continues to accumulate even after the pulse has decayed.
By the end of the observation interval, the two interferometric traces approach an antiphase relation, corresponding to an accumulated phase difference close to~$\pi$.
Despite this pronounced phase shift, the oscillation amplitudes remain comparable in the controlled and reference measurements.

To capture the intermediate-coherence regime qualitatively, we combine the two reduced descriptions of multimode polariton-condensate dynamics introduced in the preceding section.
The coherent-field description reproduces the multimode beating between states with different eigenfrequencies, but by construction represents a fully coherent field and therefore cannot account for the reduced fringe visibility observed experimentally.
In the two-particle description, the construction of the reduced correlation matrices excludes cross correlations between the trap-associated subspaces, so that its contribution contains no mutual interference between the fields associated with the two traps.
Combining the two contributions therefore provides a phenomenological description of the experimentally observed intermediate fringe visibility.
We write the calculated interferogram as
\begin{equation}
J(t,\mathbf{r};\tau)
= (1-\eta)\, J_{\mathrm{2p}}(t,\mathbf{r};\tau)
+ \eta \,J_{\mathrm{coh}}(t,\mathbf{r};\tau),
\label{EqIntermediateCoherenceMixing}
\end{equation}
where $\eta$ determines the relative weight of the coherent contribution.

%FFFFFFFFFFFFFFFFFFFFFFFF
%IIIIIIIIIIIIIIIIIIIIIIII
%GGGGGGGGGGGGGGGGGGGGGGGG
\begin{figure*}
\includegraphics[width=\textwidth]{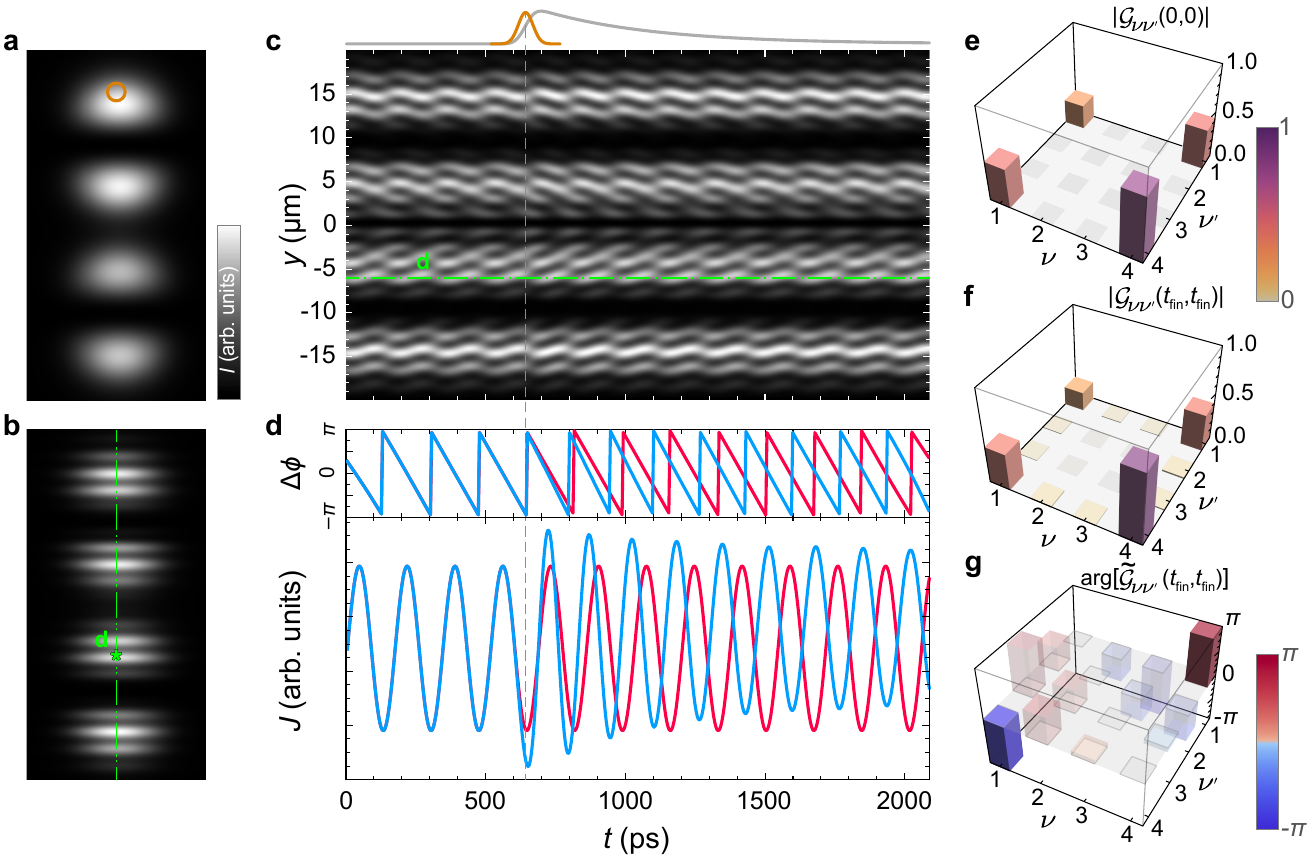}
\caption{\label{FIG_SimulHalfQuant}
\textbf{Calculated phase dynamics in the intermediate-coherence regime.}
\textbf{a}, Time-integrated condensate density.
The orange circle marks the position of the spatially localized control pulse.
\textbf{b}, Interferogram obtained by superimposing the condensate field with its copy reflected about the horizontal axis.
\textbf{c}, Space--time evolution of the interferogram along the section indicated by the green dash-dotted line in \textbf{b}, in the presence of the control pulse.
The grey curve above the panel shows schematically the pulse-induced correction to the trap potential, together with the control pulse in orange.
The vertical orange dashed line marks the arrival time of the control pulse.
\textbf{d}, Interferometric signal $J(t)$ at the position marked by the green star in \textbf{b} and the green dash-dotted line in \textbf{c} (lower panel), together with the phase $\Delta\phi(t)$ of the corresponding oscillations (upper panel).
Red and blue curves correspond to the evolutions without and with the control pulse, respectively.
\textbf{e} and \textbf{f}, Absolute values of the coherent-field correlation matrix elements at the beginning, $|\mathcal{G}_{\nu\nu'}(0,0)|$ (\textbf{e}), and at the end of the evolution, $|\mathcal{G}_{\nu\nu'}(t_{\mathrm{fin}},t_{\mathrm{fin}})|$ (\textbf{f}).
\textbf{g}, Pulse-induced phase shifts $\arg[\widetilde{\mathcal{G}}_{\nu\nu'}(t_{\mathrm{fin}},t_{\mathrm{fin}})]$ of the coherent-field correlation matrix elements at the end of the evolution, evaluated relative to the reference evolution without the control pulse.
Bars corresponding to matrix elements with small absolute values are shown with reduced opacity.
}
\end{figure*}

For this trap configuration, the coherent-field component is initialized as a superposition of the states $\Phi_{1}(\mathbf{r})$ and $\Phi_{4}(\mathbf{r})$, whereas the two-particle component is initialized in the state $\Psi_{4}(\mathbf{r}_{A},\mathbf{r}_{B})$.
The initial conditions are chosen so that both components reproduce the vertically oriented dipolar character inferred from the experimentally measured PL.
The frequency difference between $\Phi_{1}(\mathbf{r})$ and $\Phi_{4}(\mathbf{r})$ gives rise to the temporal beating, while the two-particle component reduces the fringe visibility in the interferogram.

In the coherent-field description, the model Hamiltonian is projected onto the four retained double-trap eigenstates $\Phi_{\nu}(\mathbf{r})$, yielding the evolution of their complex amplitudes $u_{\nu}(t)$.
In the two-particle description, the Hamiltonian is projected onto the four product states $\psi_{a}^{(A)}(\mathbf{r}_{A})\psi_{b}^{(B)}(\mathbf{r}_{B})$, yielding the evolution of the coefficients $C_{\mu}(t)$, where $\mu=2(a-1)+b$.
The synchronized pulse train is not included explicitly in the calculation.
Its role in preparing the multimode state is incorporated through the initial values of $u_{\nu}(0)$ and $C_{\mu}(0)$.
The explicit equations of motion and the initial conditions are given in Methods.

The control pulse is included in the model as a spatially localized, time-dependent correction to the stationary trap potential.
This correction originates from a pulse-induced perturbation of the exciton reservoir and reflects its buildup during the optical pulse and subsequent decay.
The resulting change in the potential modifies the projected Hamiltonian, transiently shifting the frequencies of the participating states and thereby changing the evolution of their complex amplitudes.
The explicit form of this correction is given in Methods.

The calculated observables are shown in Fig.~\ref{FIG_SimulHalfQuant}.
The time-integrated density and representative interferogram in panels \textbf{a} and \textbf{b} reproduce the vertically oriented dipolar character imposed through the initial conditions.
The space--time evolution of the interferogram in panel \textbf{c} captures the principal features observed in the experiment, including the well-defined inclined fringes and the reduced fringe visibility characteristic of the intermediate-coherence configuration.

The interferometric signal $J(t)$ calculated at the marked position and the phase of the corresponding oscillations are shown in Fig.~\ref{FIG_SimulHalfQuant}d.
Their evolution reproduces the experimental behaviour, with the reference and controlled traces coinciding before the control pulse and subsequently developing a phase offset that approaches $\pi$ by the end of the calculated interval, while retaining comparable oscillation amplitudes.

The pulse-induced correction to the trap potential is shown schematically by the grey curve above Fig.~\ref{FIG_SimulHalfQuant}c, together with the orange curve representing the control pulse.
The delayed maximum and slow decay of this correction reflect the buildup and relaxation of the pulse-induced perturbation of the exciton reservoir.
Such a response accounts for the delayed onset of the phase shift between the interferometric oscillations with and without the control pulse and for its continued accumulation after the optical pulse has vanished.

To examine how the control pulse affects the individual mode components, Figs.~\ref{FIG_SimulHalfQuant}e and \ref{FIG_SimulHalfQuant}f show the absolute values of the coherent-field correlation matrix $\mathcal{G}(t,t)$ at the beginning, $t=0$, and at the end, $t=t_{\mathrm{fin}}$, of the calculated evolution.
The corresponding pulse-induced phase shifts of the matrix elements at $t_{\mathrm{fin}}$ are shown in Fig.~\ref{FIG_SimulHalfQuant}g.
For each element, the phase shift is evaluated relative to the reference evolution without the control pulse as
$\arg[\tilde{\mathcal{G}}(t,t)] = \arg[\mathcal{G}(t,t)\mathcal{G}^{*}_{\mathrm{ref}}(t,t)]$,
with the result taken modulo $2\pi$.

The matrix elements associated with the initially populated states $\Phi_{1}(\mathbf{r})$ and $\Phi_{4}(\mathbf{r})$ remain dominant throughout the evolution and change only slightly in magnitude.
The off-diagonal coherence between these states acquires a phase shift close to $\pi$.
Weak matrix elements involving the other states of the reduced basis also appear, but their magnitudes remain much smaller than those associated with $\Phi_{1}(\mathbf{r})$ and $\Phi_{4}(\mathbf{r})$.
We return to these pulse-induced changes in the correlation-matrix elements in the Discussion, where they are interpreted in terms of an effective phase operation acting within the four-mode subspace.

%SSSSSSSSSSSSSSSSSS
%SSSSSSSSSSSSSSSSSS
%SSSSSSSSSSSSSSSSSS
%EEEEEEEEEEEEEEEEEE
%CCCCCCCCCCCCCCCCCC
\subsubsection*{Phase control in the high-coherence regime}

\begin{figure*}
\includegraphics[width=0.75\textwidth]{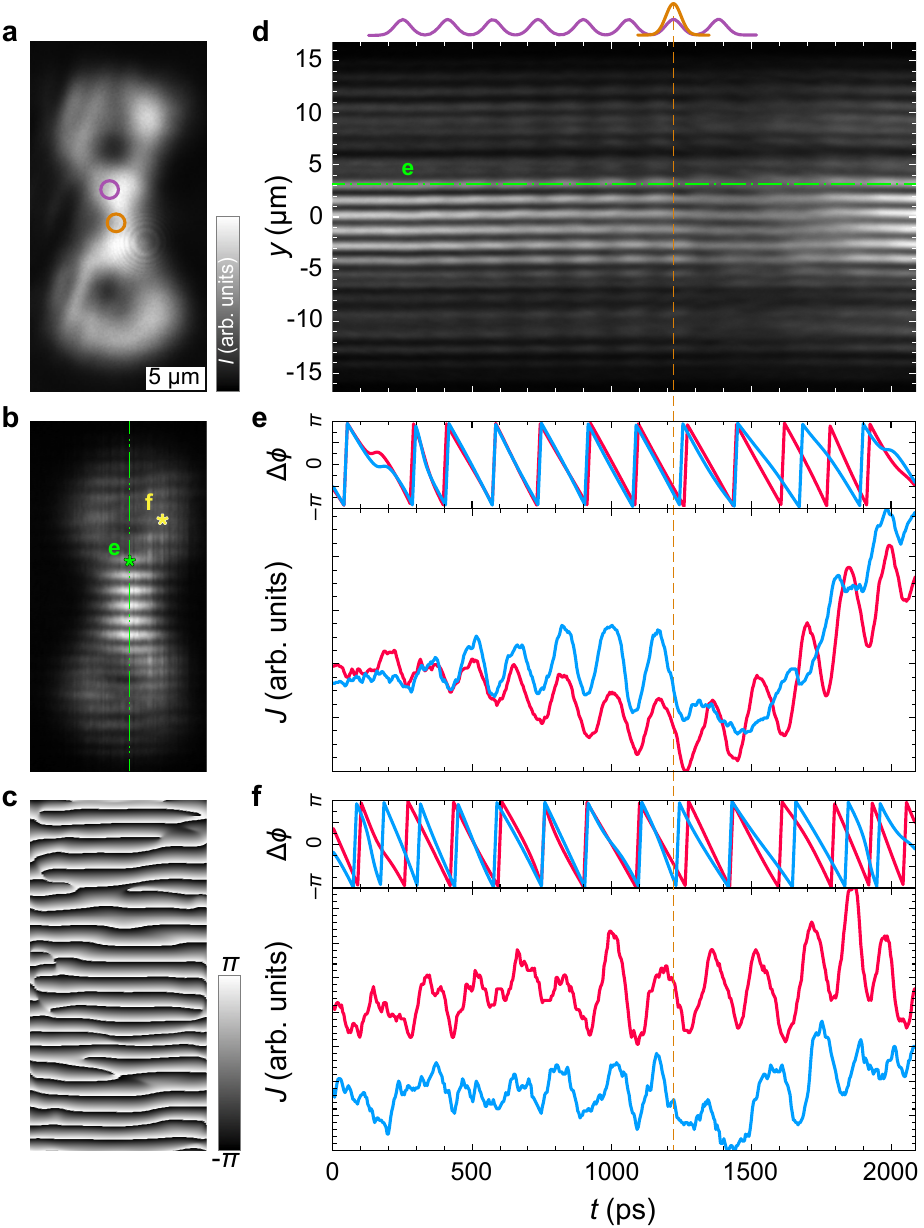}
\caption{\label{FIG_Coher}
\textbf{Pulse-induced phase control in the high-coherence regime.}
\textbf{a}, Time-integrated real-space PL of the polariton condensates.
The purple and orange circles mark the synchronized pulse train and the spatially localized control pulse, respectively.
\textbf{b}, Representative interferogram obtained by superimposing two real-space images of the condensate PL, one reflected about the horizontal axis.
\textbf{c}, Spatial phase distribution extracted from the interferogram in \textbf{b}.
\textbf{d}, Space--time evolution of the interferogram along the section indicated by the green dash-dotted line in \textbf{b}, measured in the presence of the control pulse.
The purple and orange waveforms above the panel indicate the synchronized pulse train and the control pulse, respectively.
The vertical orange dashed line marks the arrival time of the control pulse.
\textbf{e}, Interferometric signal $J(t)$ at the position marked by the green star in \textbf{b} and the green dash-dotted line in \textbf{d} (lower panel), together with the phase $\Delta\phi(t)$ of the corresponding oscillations (upper panel).
\textbf{f}, Corresponding interferometric signal and phase evolution at the position marked by the yellow star in \textbf{b}.
In \textbf{e} and \textbf{f}, red curves correspond to the reference measurement without the control pulse, whereas blue curves correspond to the otherwise identical measurements with the control pulse.
}
\end{figure*}

We next consider a trap configuration selected from the high-coherence plateau in Fig.~\ref{FIG_Coherence}c.
The time-integrated PL in Fig.~\ref{FIG_Coher}a shows intense emission throughout the region connecting the two traps, reflecting the strong extension of the condensate field across the wide intertrap channel.
The observed spatial distribution indicates that states with markedly different spatial structures contribute to the condensate dynamics.
The bright emission filling the channel points to the participation of the lowest double-trap state $\Phi_{1}(\mathbf{r})$, whose vertically oriented dipolar components strongly overlap in this region, as illustrated in the leftmost panel of Fig.~\ref{FIG_Eigs}c.
At the same time, the transverse structure within the traps indicates the contribution of another state retaining a predominantly horizontal dipolar character.

Further evidence for this state composition is provided by the representative interferogram in Fig.~\ref{FIG_Coher}b.
The interference fringes exhibit weak fork-like dislocations.
These dislocations become considerably more apparent in the spatial phase distribution extracted from the interferogram, as shown in Fig.~\ref{FIG_Coher}c.
The phase singularities within the traps are consistent with interference between states derived from vertically and horizontally oriented dipolar modes.
The measured beating frequency provides an additional constraint on the identification of the participating states, which we use below in constructing the coherent-field description.

%FFFFFFFFFFFFFFFFFFFFFFFF
%IIIIIIIIIIIIIIIIIIIIIIII
%GGGGGGGGGGGGGGGGGGGGGGGG

%FFFFFFFFFFFFFFFFFFFFFFFF
%IIIIIIIIIIIIIIIIIIIIIIII
%GGGGGGGGGGGGGGGGGGGGGGGG
\begin{figure*}
\includegraphics[width=\textwidth]{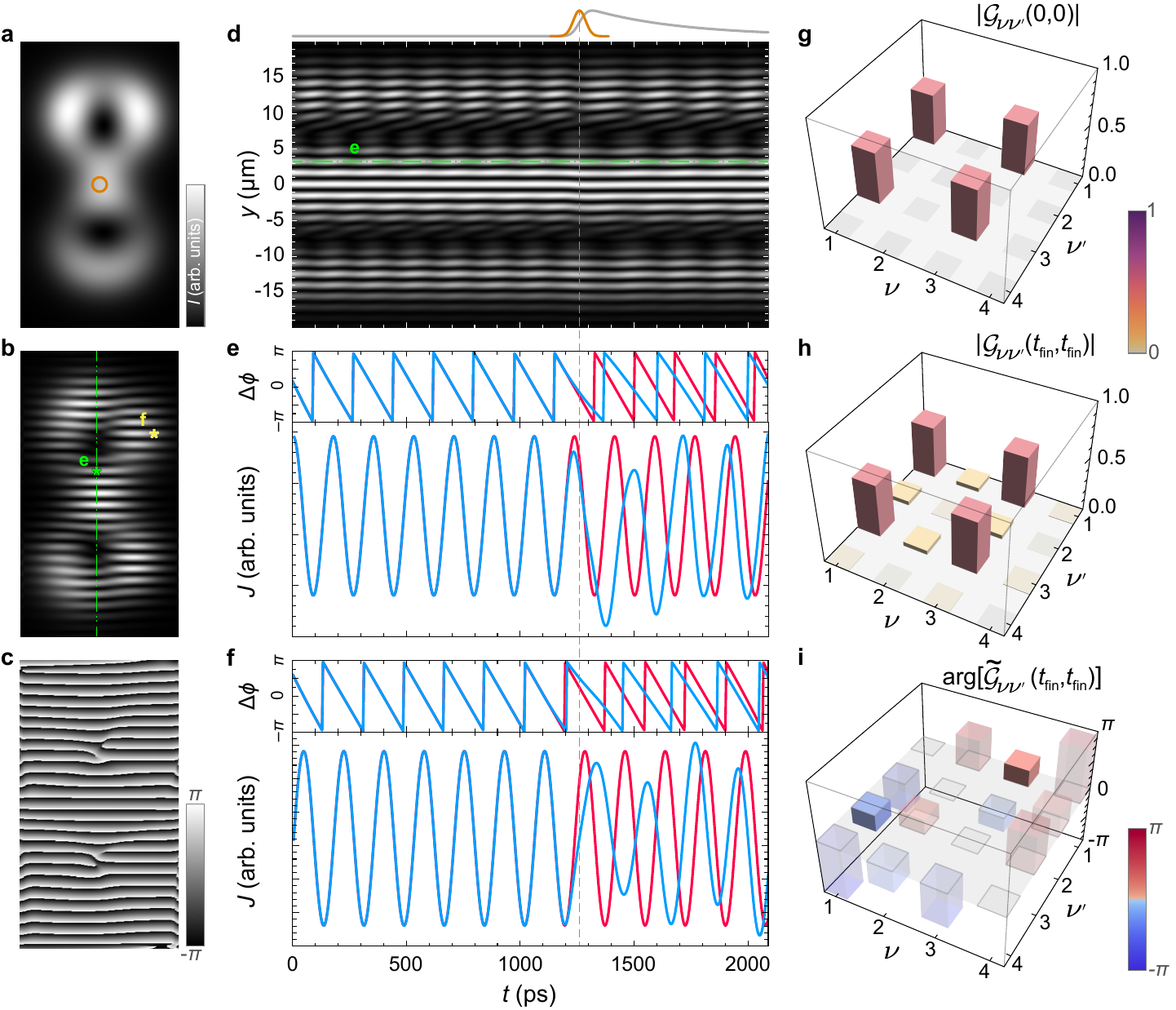}
\caption{\label{FIG_SimulCoher}
\textbf{Calculated phase dynamics in the high-coherence regime.}
\textbf{a}, Time-integrated condensate density.
The orange circle marks the position of the spatially localized control pulse.
\textbf{b}, Interferogram obtained by superimposing the condensate field with its copy reflected about the horizontal axis.
\textbf{c}, Spatial phase distribution extracted from the interferogram in \textbf{b}.
\textbf{d}, Space--time evolution of the interferogram along the section indicated by the green dash-dotted line in \textbf{b}, in the presence of the control pulse.
The grey curve above the panel shows schematically the pulse-induced correction to the trap potential, together with the control pulse in orange.
The vertical orange dashed line marks the arrival time of the control pulse.
\textbf{e}, Interferometric signal $J(t)$ at the position marked by the green star in \textbf{b} and the green dash-dotted line in \textbf{d} (lower panel), together with the phase $\Delta\phi(t)$ of the corresponding oscillations (upper panel).
\textbf{f}, Corresponding interferometric signal and phase evolution at the position marked by the yellow star in \textbf{b}.
In \textbf{e} and \textbf{f}, red and blue curves correspond to the evolutions without and with the control pulse, respectively.
\textbf{g} and \textbf{h}, Absolute values of the coherent-field correlation matrix elements at the beginning, $|\mathcal{G}_{\nu\nu'}(0,0)|$ (\textbf{g}), and at the end of the evolution, $|\mathcal{G}_{\nu\nu'}(t_{\mathrm{fin}},t_{\mathrm{fin}})|$ (\textbf{h}).
\textbf{i}, Pulse-induced phase shifts $\arg[\widetilde{\mathcal{G}}_{\nu\nu'}(t_{\mathrm{fin}},t_{\mathrm{fin}})]$ of the coherent-field correlation matrix elements at the end of the evolution, evaluated relative to the reference evolution without the control pulse.
Bars corresponding to matrix elements with small absolute values are shown with reduced opacity.
}
\end{figure*}

The temporal evolution of the interferogram is followed along the $y$-directed section indicated by the green dash-dotted line in Fig.~\ref{FIG_Coher}b.
The section is displaced slightly from the symmetry axis of the structure, allowing it to sample both the intense emission in the intertrap channel and the weaker laterally distributed emission within the traps.
The resulting space--time map is shown in Fig.~\ref{FIG_Coher}d.
As in the intermediate-coherence configuration, the periodic structure along $y$ is imposed by the carrier wave-vector component $q_y$.
Here, however, the temporal oscillations appear primarily as a modulation of the nearly horizontal fringes along $t$.
The comparatively moderate modulation depth reflects the strongly unequal local contributions of the states participating in the oscillations within the selected section.

The synchronized pulse train and the control pulse are indicated schematically by the purple and orange waveforms above Fig.~\ref{FIG_Coher}d, respectively.
The effect of the control pulse is examined at two positions where the states participating in the oscillations contribute with different relative weights.
The first position, marked by the green star in Fig.~\ref{FIG_Coher}b, lies within the section displayed in Fig.~\ref{FIG_Coher}d and samples the intense emission in the channel.
The second position, marked by the yellow star, is displaced towards a lateral emission maximum within the trap, where the overall signal is weaker but the temporal modulation remains clearly visible.

Figures~\ref{FIG_Coher}e and \ref{FIG_Coher}f show the interferometric signals $J(t)$ and the phases $\Delta\phi(t)$ of the corresponding oscillations at the two selected positions.
At both positions, the phase evolutions with and without the control pulse nearly coincide before its arrival, after which a phase shift develops.
Because the control pulse arrives comparatively late within the observation window, this phase shift is still evolving at the end of the measurement and does not reach a stationary value within the recorded interval.

The high mutual coherence of this configuration allows its dynamics to be described using the coherent-field model alone.
The strong emission in the intertrap channel points to $\Phi_{1}(\mathbf{r})$ as one of the participating states, while among the two horizontal-dipolar states of the reduced basis, $\Phi_{3}(\mathbf{r})$ is selected by the measured beating frequency, which agrees with the eigenfrequency difference $\omega_{3}-\omega_{1}$.
The calculated observables are presented in Fig.~\ref{FIG_SimulCoher}.
The time-integrated density in Fig.~\ref{FIG_SimulCoher}a reproduces the enhanced condensate density in the channel together with the lateral structure associated with the horizontal-dipolar state.
The interferogram and spatial phase distribution in Figs.~\ref{FIG_SimulCoher}b and \ref{FIG_SimulCoher}c reproduce the fringe dislocations and phase singularities observed experimentally.
The space--time map in Fig.~\ref{FIG_SimulCoher}d likewise captures the nearly horizontal carrier fringes modulated by the temporal beating between $\Phi_{1}(\mathbf{r})$ and $\Phi_{3}(\mathbf{r})$.

The calculated interferometric signals $J(t)$ and the phases $\Delta\phi(t)$ of the corresponding oscillations at the two marked positions are shown in Figs.~\ref{FIG_SimulCoher}e and \ref{FIG_SimulCoher}f.
At both positions, the calculation reproduces the experimentally observed development of the phase shift following the control pulse.
To examine how the control pulse affects the individual mode components, Figs.~\ref{FIG_SimulCoher}g and \ref{FIG_SimulCoher}h show the absolute values of the correlation matrix $\mathcal{G}(t,t)$ at $t=0$ and $t=t_{\mathrm{fin}}$, respectively.
The corresponding pulse-induced phase shifts of the matrix elements at $t_{\mathrm{fin}}$ are shown in Fig.~\ref{FIG_SimulCoher}i.
The matrix elements associated with the initially populated states $\Phi_{1}(\mathbf{r})$ and $\Phi_{3}(\mathbf{r})$ remain dominant, while only weak elements involving the other states of the reduced basis appear.
The off-diagonal elements $\mathcal{G}_{13}$ and $\mathcal{G}_{31}$ acquire pronounced phase shifts of opposite sign.

%SSSSSSSSSSSSSSSSSS
%SSSSSSSSSSSSSSSSSS
%EEEEEEEEEEEEEEEEEE
%CCCCCCCCCCCCCCCCCC
\section*{Discussion}

Control over multimode polariton-condensate dynamics in spatially structured trapping potentials requires both the establishment of coherence across the structure and the ability to manipulate the phase relations among the participating states.
The width of the channel connecting the two optically induced traps controls their mutual coherence, with the sharp rise in the zero-delay first-order coherence marking the onset of a regime in which reproducible phase correlations extend across the coupled structure.
Importantly, the development of mutual coherence does not eliminate the multimode character of the condensate dynamics, as evidenced by the intensity oscillations arising from interference between several populated states.
The phase relations among the mutually coherent states are therefore well-defined dynamical degrees of freedom that can be modified by a spatially localized optical control pulse.
The channel geometry thus establishes the coherence regime, whereas the control pulse transforms the phase relations within an already formed multimode state without substantially changing its populated-state composition.

The observation of phase control in two distinct modal configurations indicates that the effect is not tied to a particular pair of states.
In general, the reservoir perturbation induced by the control pulse can modify both the phases and the amplitudes of the participating states.
The balance between these effects is determined by the spatial overlap of the perturbation with the participating states and by its strength and duration.
Under the conditions realized in our experiments, the system approaches a phase-selective regime in which the control pulse acts predominantly on the phase relations, producing a pronounced phase shift between the initially populated states while leaving their amplitudes nearly unchanged and the admixture of additional states weak.
Because the reservoir perturbation builds up and relaxes more slowly than the optical pulse, the resulting phase shift develops with a finite delay and continues to evolve after the pulse has vanished.

Phase control is already available in the intermediate-coherence regime, before mutual coherence reaches its high-coherence plateau.
In this regime, the mixed interferogram introduced in Eq.~\ref{EqIntermediateCoherenceMixing} reproduces the oscillatory phase dynamics and reduced fringe visibility by combining a coherent-field contribution with one that contains no intertrap correlations.

In both coherence regimes, the localized control pulse induces a state-dependent phase transformation of the coherent multimode contribution.
This action can be represented by an effective multimode phase operator defined in the reduced basis~$\{\Phi_{\nu}(\mathbf{r})\}_{\nu=1}^{4}$,
\begin{equation}
\mathcal{S} = \operatorname{diag}
\left(
e^{i\phi_{1}}, e^{i\phi_{2}}, e^{i\phi_{3}}, e^{i\phi_{4}}
\right),
\label{EqDiscussionPhaseOperator}
\end{equation}
where $\phi_{\nu}$ is the additional phase acquired by the state $\Phi_{\nu}(\mathbf{r})$ relative to the reference evolution without the control pulse.
The corresponding transformation of the coherence matrix is
$\mathcal{G}_{\mathrm{tar}} = \mathcal{S}^{\dagger} \mathcal{G}^{(\mathrm{ref})} \mathcal{S}$,
so that each off-diagonal element acquires the phase shift
$\delta\phi_{\nu\nu'}=\phi_{\nu'}-\phi_{\nu}$.
Only these phase differences are physically relevant, since a common phase leaves the coherence matrix unchanged.
When two states dominate the condensate field, their phase difference is directly reflected in the phase of the observed intensity oscillations.

The mode-resolved action of the effective phase operator is directly reflected in the calculated coherence matrices.
In the intermediate-coherence configuration, the coherent-field contribution is initialized as a superposition of $\Phi_{1}(\mathbf{r})$ and $\Phi_{4}(\mathbf{r})$, and the matrix elements associated with these states remain dominant throughout the evolution, as seen from their absolute values in Figs.~\ref{FIG_SimulHalfQuant}e and \ref{FIG_SimulHalfQuant}f.
The control pulse introduces an additional phase difference close to $\pi$ between these states, which appears in Fig.~\ref{FIG_SimulHalfQuant}g as opposite phase shifts of the off-diagonal elements $\mathcal{G}_{14}$ and $\mathcal{G}_{41}$.
In the high-coherence configuration, the condensate field is initialized in $\Phi_{1}(\mathbf{r})$ and $\Phi_{3}(\mathbf{r})$, which likewise remain the dominant states according to the absolute values of the matrix elements in Figs.~\ref{FIG_SimulCoher}g and \ref{FIG_SimulCoher}h.
The corresponding phase transformation appears in Fig.~\ref{FIG_SimulCoher}i as opposite phase shifts of $\mathcal{G}_{13}$ and $\mathcal{G}_{31}$.
The two configurations therefore show that the same form of effective phase operation applies to different pairs of states despite the change in the underlying spatial modes with trap geometry.

To quantify how closely the calculated control-pulse action approaches the effective phase transformation, we use the normalized overlap of the two coherence matrices as a fidelity measure,
\begin{equation}
\mathcal{F}
=
\frac{
\operatorname{Tr}
\left(
\mathcal{G}_{\mathrm{ctrl}}
\mathcal{G}_{\mathrm{tar}}
\right)
}{
\operatorname{Tr}
\left(
\mathcal{G}_{\mathrm{ctrl}}
\right)
\operatorname{Tr}
\left(
\mathcal{G}_{\mathrm{tar}}
\right)
}.
\label{EqDiscussionPhaseFidelity}
\end{equation}
Here, $\mathcal{G}_{\mathrm{ctrl}}\equiv\mathcal{G}^{(\mathrm{ctrl})}(t_{\mathrm{fin}},t_{\mathrm{fin}})$ is the coherence matrix at the end of the calculated evolution with the control pulse, while
$\mathcal{G}_{\mathrm{tar}}\equiv\mathcal{S}^{\dagger}\mathcal{G}^{(\mathrm{ref})}(t_{\mathrm{fin}},t_{\mathrm{fin}})\mathcal{S}$
is the target matrix obtained by applying the effective phase operator to the reference coherence matrix at the same final time.
With the phases $\phi_{\nu}$ chosen from the calculated pulse-induced shifts, the fidelity exceeds $0.995$ in both coherence configurations.
This close agreement shows that the deviations from a pure phase transformation remain small despite the accompanying changes in amplitudes and the weak admixture of additional states.
The pulse position, fluence, duration, and arrival time provide a natural set of control parameters for extending the demonstrated operation towards programmable multimode phase transformations.

{\color{blue}

}%color

%SSSSSSSSSSSSSSSSSS
%SSSSSSSSSSSSSSSSSS
%EEEEEEEEEEEEEEEEEE
%CCCCCCCCCCCCCCCCCC
\subsection*{Methods}

%SSSSSSSSSSSSSSSSSS
%SSSSSSSSSSSSSSSSSS
%EEEEEEEEEEEEEEEEEE
%CCCCCCCCCCCCCCCCCC
\subsection*{Sample details}

The sample is a planar optical microcavity with top and bottom distributed Bragg reflectors comprising 40 and 45 AlAs/GaAlAs $\lambda/4$ layer pairs, respectively.
The $3\lambda/2$ AlGaAs cavity contains three sets of four GaAs quantum wells positioned at the antinodes of the cavity mode. 
The structure has a quality factor of approximately $1.6\times10^4$ and a vacuum Rabi splitting of 5~meV.
A thickness wedge in the cavity layer allows the exciton-photon detuning to be selected through the excitation position, with all measurements performed at a detuning of $-0.5$~meV.
The sample was maintained at 6~K in a cold-finger cryostat.

%SSSSSSSSSSSSSSSSSS
%SSSSSSSSSSSSSSSSSS
%EEEEEEEEEEEEEEEEEE
%CCCCCCCCCCCCCCCCCC
\subsection*{Optical excitation and detection}

The optical excitation and detection scheme is shown in Fig.~\ref{FIG_Scheme}b.
The polariton condensates were populated nonresonantly using a CW single-mode semiconductor laser tuned above the lower-polariton branch.
A digital micromirror device acting as a spatial light modulator (DMD SLM) shaped the pump beam into two adjacent ring profiles forming the optically induced traps, whose relative positions were varied to control the channel width $L$.
The excitation power was monitored with a power meter (PM), while unused beams were terminated on black-body absorbers (BB).

To enhance the polariton-density oscillations, an $80$~fs pulse from a mode-locked Ti:sapphire laser was divided into two branches.
In one branch, successive reflections from four parallel glass plates of equal thickness generated a train of eight pulses, with the interval between neighboring pulses adjusted to match the oscillation period for each trap configuration.
The reflected pulse train was separated from the incident beam using a half-wave plate ($\lambda/2$) and a polarizing beam splitter (PBS) and directed toward the sample.
The second branch provided the control pulse, whose arrival time was adjusted with a motorized delay line and whose position on the sample was controlled independently.

The emission from the sample was collected by a microscope objective (MO), with residual excitation light suppressed by an FELH spectral filter, and directed to imaging, interferometric, or time-resolved detection channels.
The real-space condensate PL was recorded with a CMOS camera.
For interferometric measurements, the emission was directed to the MZI, where a Dove prism in one arm inverted the real-space image according to $y\rightarrow -y$, allowing the two trap regions to be superimposed upon recombination.
A small relative angle between the two arms introduced carrier fringes, while their temporal delay $\tau$ was varied for measurements of $g^{(1)}(\tau,\mathbf{r})$.
The interferograms were recorded with a CCD camera, while the temporal evolution of the condensate PL was measured with a streak camera coupled to a monochromator (MC) and synchronized with the excitation sequence.

%SSSSSSSSSSSSSSSSSS
%SSSSSSSSSSSSSSSSSS
%EEEEEEEEEEEEEEEEEE
%CCCCCCCCCCCCCCCCCC
\subsection*{Extraction of mutual first-order coherence}

The mutual first-order coherence was extracted from real-space interferograms recorded with the MZI and accumulated over the camera exposure time at each optical delay $\tau$.
Owing to the image inversion in one interferometer arm, the fields $E_1(t,\mathbf{r})$ and $E_2(t,\mathbf{r})$ arriving through the two arms at the same detector point $\mathbf{r}=(x,y)$ originated from $\mathbf{r}$ and $\bar{\mathbf{r}}=(x,-y)$ in the condensate image, respectively.
A small relative angle between the recombined images introduced the carrier component $q_{y}$.
The time-averaged interferogram can therefore be written as
\begin{equation}
J(\tau,\mathbf{r}) =
I_1(\mathbf{r}) + I_2(\mathbf{r})
- 2\operatorname{Re}\left[\mathcal{J}(\tau,\mathbf{r})e^{iq_{y}y}\right],
\end{equation}
where $\mathcal{J}(\tau,\mathbf{r})=\left\langle E_1^{*}(\mathbf{r},t)E_2(\mathbf{r},t+\tau)\right\rangle_t$ and $I_{1,2}(\mathbf{r})=\left\langle|E_{1,2}(\mathbf{r},t)|^2\right\rangle_t$ are the field correlation and time-averaged intensities in the two interferometer arms, respectively.
Here, $\langle\ldots\rangle_t$ denotes averaging over the interferogram accumulation time.
The intensity distributions $I_1(\mathbf{r})$ and $I_2(\mathbf{r})$ were measured separately by blocking the opposite interferometer arm.

Each interferogram was Fourier transformed in the spatial coordinates, and one sideband centred at $q_{y}$ was isolated, shifted to the origin, and transformed back to real space.
The modulus of the reconstructed complex amplitude gives $|\mathcal{J}(\tau,\mathbf{r})|$, from which the spatially resolved degree of mutual first-order coherence was calculated as
\begin{equation}
g^{(1)}(\tau,\mathbf{r}) = |\mathcal{J}(\tau,\mathbf{r})| \left/
\sqrt{I_1(\mathbf{r})I_2(\mathbf{r})} \right. .
\end{equation}

For each channel width $L$, the delay dependence was evaluated at the point $\mathbf{r}^{\ast}_L$ of maximum intensity overlap between the superimposed images.
The dependencies presented in Fig.~\ref{FIG_Coherence}a therefore correspond to $g^{(1)}(\tau)\equiv g^{(1)}(\tau,\mathbf{r}^{\ast}_L)$.

%SSSSSSSSSSSSSSSSSS
%SSSSSSSSSSSSSSSSSS
%EEEEEEEEEEEEEEEEEE
%CCCCCCCCCCCCCCCCCC
\subsection*{Construction of model potentials}

In the experiment, localization of the polariton condensate field within the optically induced traps is sustained by the driven-dissipative nature of the polariton system.
To obtain localized basis states from conservative stationary eigenvalue problems, we model each trap by a smooth finite-depth potential well that approaches a constant value $V_{0}$ outside the trapping region.

The construction of the model potential follows the same general principle as the formation of the smooth experimental excitation profile from a binary spatial-light-modulator pattern.
For each isolated trap $j=A,B$, centered at $(0,y_{\mathrm{c}j})$, we define the step-like potential profile $V^{\mathrm{step}}_j(\mathbf{r}) = 0$ for $(x/s)^{2}+[s(y-y_{\mathrm{c}j})]^{2}\leq R_{j}^{2}$ and $V^{\mathrm{step}}_j(\mathbf{r}) = 1$ otherwise, where $R_{j}$ and $s$ determine the trap size and ellipticity, respectively.
Its sharp boundary is smoothed by convolution with a normalized two-dimensional Gaussian kernel $K_{w}(\mathbf{r})$, whose standard deviation $w$ sets the width of the smoothed boundary, yielding the isolated-trap potential $V_{j}(\mathbf{r}) = V_{0}\left[K_{w}\ast V_{j}^{\mathrm{step}}\right](\mathbf{r})$.
%$K_{w}(\mathbf{r})=(2\pi w^{2})^{-1}\exp[-|\mathbf{r}|^{2}/(2w^{2})]$

The step-like double-trap profile is constructed as $W^{\mathrm{step}}(\mathbf{r}) = V_{A}^{\mathrm{step}}(\mathbf{r})V_{B}^{\mathrm{step}}(\mathbf{r})$, and Gaussian smoothing yields
$ W(\mathbf{r}) = V_{0}\left[K_{w}\ast W^{\mathrm{step}}\right](\mathbf{r})$.
The overlap of the two elliptic regions forms the channel between the traps.
For fixed trap geometry, its width $L$ is uniquely determined by the center-to-center separation $d=|y_{\mathrm{c}B}-y_{\mathrm{c}A}|$.
The parameters $V_{0}$, $s$, and $w$ are common to both traps, whereas their radii $R_{A}$ and $R_{B}$ may differ.

The action of the control pulse is included as a localized time-dependent perturbation $\delta V(t,\mathbf{r})$ of the reservoir-induced effective potential.
Assuming that the reservoir perturbation relaxes exponentially with decay rate $\gamma_{\mathrm{R}}$, its response to the control pulse is represented by the causal temporal convolution
$ \delta V(t,\mathbf{r}) = \int_{-\infty}^{t} \exp[{-\gamma_{\mathrm{R}}(t-t')}] P_{\mathrm{c}}(t',\mathbf{r}) \,dt'$,
where $P_{\mathrm{c}}(t,\mathbf{r})$ is a phenomenological pump term with the spatial and temporal profiles following those of the optical control pulse.
Its amplitude determines the strength of the induced potential perturbation and is treated as a model parameter.

%\begin{equation}
%\partial_t \delta V(t,\mathbf{r}) 
%= P_{\mathrm{c}}(t,\mathbf{r}) - \gamma_{\mathrm{R}}\delta V(t,\mathbf{r}),
%\end{equation}

%SSSSSSSSSSSSSSSSSS
%SSSSSSSSSSSSSSSSSS
%EEEEEEEEEEEEEEEEEE
%CCCCCCCCCCCCCCCCCC
\subsection*{Amplitude dynamics in the two-particle model}

In the two-particle model, the dynamics of the polariton condensates in the coupled-trap system is described by the two-coordinate wavefunction $\Psi(t,\mathbf{r}_{A},\mathbf{r}_{B})$, which obeys the effective Schr\"{o}dinger equation
\begin{equation}
i\hbar\,\partial_t \Psi(t,\mathbf{r}_{A},\mathbf{r}_{B})
=
\sum_{j=A,B}
\hat{H}_{j}(t,\mathbf{r}_{j})
\Psi(t,\mathbf{r}_{A},\mathbf{r}_{B}),
\label{EqGeneralTwoCoordEvolutionPaper}
\end{equation}
where the single-coordinate Hamiltonians are
$\hat{H}_{j}(t,\mathbf{r}_{j}) = \hat{H}_0(\mathbf{r}_{j}) + \delta V(t,\mathbf{r}_{j})$,
with both terms evaluated at the corresponding coordinate $\mathbf{r}_{j}$.
Projecting Eq.~\eqref{EqGeneralTwoCoordEvolutionPaper} onto the four-mode product basis introduced in Eq.~\eqref{EqTwoParticleWaveFunction} yields the evolution equations for the basis-state amplitudes $C_{\mu}(t)$.
Using the two-index notation $\mu\leftrightarrow(a,b)$, they take the form
\begin{equation}
i\hbar\,\dot{C}_{ab}(t)
= \sum_{a'=1}^{2} h^{(A)}_{aa'}(t)C_{a'b}(t)
+ \sum_{b'=1}^{2} h^{(B)}_{bb'}(t)C_{ab'}(t),
\label{EqAmplitudeEvolutionFactorized}
\end{equation}
where $h^{(j)}_{nn'}(t) = \int \psi_{n}^{(j)*}(\mathbf{r}) \hat{H}_{j} (t,\mathbf{r}) \psi_{n'}^{(j)}(\mathbf{r}) \,d\mathbf{r}$, $ n, n'=1,2$.
The basis functions are kept fixed throughout the calculation, while the control pulse enters through the time dependence of the projected matrix elements.

%SSSSSSSSSSSSSSSSSS
%SSSSSSSSSSSSSSSSSS
%EEEEEEEEEEEEEEEEEE
%CCCCCCCCCCCCCCCCCC
\subsection*{Amplitude dynamics in the coherent-field model}

In the coherent-field model, the polariton condensate field $\Phi(t,\mathbf{r})$ obeys
\begin{equation}
i\hbar\,\partial_t \Phi(t,\mathbf{r}) = \hat{H}(t,\mathbf{r})\Phi(t,\mathbf{r}),
\label{EqCoherentFieldEvolution}
\end{equation}
where $\hat{H}(t,\mathbf{r})=\hat{H}_{0}(\mathbf{r})+\delta V(t,\mathbf{r})$.
Projecting Eq.~\eqref{EqCoherentFieldEvolution} onto the four eigenstates $\Phi_{\nu}(\mathbf{r})$ of the stationary double-trap potential yields the evolution equations for their amplitudes $u_{\nu}(t)$,
\begin{equation}
i\hbar\,\dot{u}_{\nu}(t)
=
\sum_{\nu'=1}^{4}
\mathcal{H}_{\nu\nu'}(t)u_{\nu'}(t),
\label{EqCoherentAmplitudeEvolution}
\end{equation}
where
$
\mathcal{H}_{\nu\nu'}(t)
=
\int
\Phi_{\nu}^{*}(\mathbf{r})
\hat{H}(t,\mathbf{r})
\Phi_{\nu'}(\mathbf{r})
\,d\mathbf{r}
$,
$\nu,\nu'=1,\ldots,4$.

%SSSSSSSSSSSSSSSSSS
%SSSSSSSSSSSSSSSSSS
%EEEEEEEEEEEEEEEEEE
%CCCCCCCCCCCCCCCCCC
\subsection*{Model parameters}

The simulations were performed using the following parameters.
The effective polariton mass is $m^{*}=3 \times10^{-5} m_{\mathrm{e}}$, where $m_{\mathrm{e}}$ is the free electron mass.
The model trap potentials were defined by the potential depth $V_{0}=0.6$~meV, the trap radii $R_{A}=8.3 \, \mu \mathrm{m}$ and $R_{B}=8.5 \, \mu \mathrm{m}$, the ellipticity  $s=0.97$, and the Gaussian smoothing width $w=1.5 \, \mu \mathrm{m}$.
%For the intermediate- and high-coherence configurations, the center-to-center separations $d$ were $18.95 \, \mu \mathrm{m}$ and $16.96 \, \mu \mathrm{m}$, respectively.

The control pulse was represented by a Gaussian term in both space and time,
$
P_{\mathrm{c}}(t,\mathbf{r})
\propto
\exp\left[
-{|\mathbf{r}-\mathbf{r}_{\mathrm{c}}|^{2}}/{w_{r}^{2}}
\right]
\exp\left[
-{(t-t_{\mathrm{c}})^{2}}/{w_{t}^{2}}
\right],
$
where $w_{r}=3~\mu\mathrm{m}$ and $w_{t}=40~\mathrm{ps}$ determine the spatial and temporal widths of the pulse, respectively.
For the intermediate- and high-coherence configurations, the pulse positions were $(0,0)~\mu\mathrm{m}$ and $(15,0)~\mu\mathrm{m}$, and the corresponding arrival times were $640$~ps and $1260$~ps, respectively.
The reservoir relaxation time was $\gamma_{\mathrm{R}}^{-1}=400$~ps.

For the intermediate-coherence calculation, the coherent-field component was initialized with amplitudes
$u_{1}(0)=0.52\exp[-0.7i\pi]$ and
$u_{4}(0)=\sqrt{1-|u_{1}(0)|^{2}}$
for the states $\Phi_{1}(\mathbf{r})$ and $\Phi_{4}(\mathbf{r})$, respectively, while the two-particle component was initialized in the state $\Psi_{4}(\mathbf{r}_{A},\mathbf{r}_{B})$ with $C_{4}(0)=1$.
For the high-coherence calculation, the initial field contained the states $\Phi_{1}(\mathbf{r})$ and $\Phi_{3}(\mathbf{r})$ with amplitudes
$u_{1}(0)=u_{3}(0)=1/\sqrt{2}$.

To reproduce the experimental interference geometry, the calculated interferograms were constructed by superimposing the calculated condensate field with its copy reflected about the horizontal axis.
For the intermediate-coherence configuration, the carrier wave-vector component was $q_{y}=3\,\mu \mathrm{m}^{-1}$, the relative image displacements were $\Delta x=0.4~\mu\mathrm{m}$ and $\Delta y=-1.6~\mu\mathrm{m}$, the optical delay between the interferometer arms was $\tau=0$, and a constant phase offset $\phi_{0}=-0.75\pi$ was introduced between the interfering fields.
For the high-coherence configuration, the corresponding parameters were $q_{y}=4\,\mu \mathrm{m}^{-1}$, $\Delta x=0.8~\mu\mathrm{m}$, $\Delta y=9.8~\mu\mathrm{m}$, $\tau=38~\mathrm{ps}$, and $\phi_{0}=0.5\pi$.

\section*{Acknowledgments}
The support of the Ministry of Science and Higher Education of the Russian Federation (No. FSMG-2026-0012),
Saint-Petersburg State University (research grant No.~125022803069-4) and
the state assignment in the field of scientific activity of the Ministry of Science and Higher Education of the Russian Federation (theme FZUN-2024-0019, state assignment of VlSU) %and the Innovation Program for Quantum Science and Technology 2023ZD0300300 
are acknowledged. 

\bibliography{twoQubitsBibl}% common bib file

\end{document}